# A note on proper curvature collineations in Bianchi type IV space-times

Ghulam Shabbir and Amjad Ali

Faculty of Engineering Sciences,

GIK Institute of Engineering Sciences and Technology,

Topi, Swabi, NWFP, Pakistan.

Email: shabbir@giki.edu.pk

**Abstract**

Curvature collineations of Bianchi type IV space-times are investigated using the rank of the $6\times 6$ Riemann matrix and direct integration technique. From the above study it follows that the Bianchi type IV space-times possesses only one case when it admits proper curvature collineations. It is shown that proper curvature collineations form an infinite dimensional vector space.

## 1. INTRODUCTION

In this paper we investigate the existance of proper curvature collineations (CCS) in Bianchi type IV space-times. Curvature collineation which preserves the curvature structure of a space-time carries significant information and plays an important role in Einstein's theory of general relativity. It is therefore important to study curvature collineations. Different approaches [1-21] were adopted to study CCS. In this paper an aproach, which is given in [5], is adopted to study proper curvature collineations in Bianchi type IV space-times by using the rank of the $6\times 6$ Rieman matrix and direct integration techinques.

Throughout $M$ represents a four dimensional, connected, Hausdorff space-time manifold with Lorentz metric $g$ of signature (-, +, +, +). The curvature tensor associated with $g_{ab}$, through the Levi-Civita connection, is denoted in component form by $R^a{}_{bcd}$. The usual covariant, partial and Lie derivatives are denoted by a semicolon, a comma and the symbol $L$, respectively. Round and square brackets denote the usual symmetrization and skew-symmetrization, respectively. Here, $M$ is assumed non-flat in the sense that the curvature tensor does not vanish over any non-empty open subset of $M$.

Any vector field $X$ on $M$ can be decomposed as

$$X_{a;b} = \frac{1}{2}h_{ab} + G_{ab} \qquad (1)$$

where $h_{ab}(=h_{ba}) = L_X g_{ab}$ is a symmetric and $G_{ab}(=-G_{ba})$ is a skew symmetric tensor on $M$. If $h_{ab;c} = 0$, $X$ is said to be *affine* and further satisfies $h_{ab} = 2cg_{ab}, c \in R$ then $X$ is said to be *homothetic* (and *Killing* if c = 0). The vector field $X$ is said to be proper affine if it is not homothetic vector field and also $X$ is said to be proper homothetic vector field if it is not Killing vector field.

A vector field $X$ on $M$ is called a curvature collineation (CC) if it satisfies [1]

$$L_X R^a{}_{bcd} = 0 \qquad (2)$$

or equivalently,

$$R^a{}_{bcd;e} X^e + R^a{}_{ecd} X^e{}_{;b} + R^a{}_{bed} X^e{}_{;c} + R^a{}_{bce} X^e{}_{;d} - R^e{}_{bcd} X^a{}_{;e} = 0.$$

The vector field $X$ is said to be proper CC if it is not affine [5] on $M$.

## 2. CLASSIFICATION OF THE RIEMANN TENSORS

In this section we will classify the Riemann tensor in terms of its rank and bivector decomposition.

The rank of the Riemann tensor is the rank of the $6 \times 6$ symmetric matrix derived in a well known way [5]. The rank of the Riemann tensor at $p$ is the rank of the linear map $f$ which maps the vector space of all bivectors $G$ at $p$ to itself and is defined by $f : G^{ab} \to R^{ab}{}_{cd} G^{cd}$. Define the subspace $N_p$ of the tangent space $T_pM$ consisting of those members $k$ of $T_pM$ which satisfy the relation

$$R_{abcd} k^d = 0 \qquad (3)$$

Then the Riemann tensor at $p$ satisfies exactly one of the following algebraic conditions [5].

**Class B**

The rank is 2 and the range of $f$ is spanned by the dual pair of non-null simple bivectors and $\dim N_p = 0$. The Riemann tensor at $p$ takes the form

$$R_{abcd} = \alpha\, G_{ab} G_{cd} + \beta\, \overset{*}{G}_{ab}\, \overset{*}{G}_{cd} \tag{4}$$

where $G$ and its dual $\overset{*}{G}$ are the (unique up to scaling) simple non-null spacelike and timelike bivectors in the range of $f$, respectively and $\alpha, \beta \in R$.

**Class C**

The rank is 2 or 3 and there exists a unique (up to scaling) solution say, $k$ of (3) (and so $\dim N_p = 1$). The Riemann tensor at $p$ takes the form

$$R_{abcd} = \sum_{i,j=1}^{3} \alpha_{ij}\, G^i{}_{ab} G^j{}_{cd} \tag{5}$$

where $\alpha_{ij} \in R$ for all $i, j$ and $G^i{}_{ab} k^b = 0$ for each of the bivectors $G^i$ which span the range of $f$.

**Class D**

Here the rank of the curvature matrix is 1. The range of the map $f$ is spanned by a single bivector $G$, say, which has to be simple because the symmetry of Riemann tensor $R_{a[bcd]} = 0$ means $G_{a[b} G_{cd]} = 0$. Then it follows from a standard result that $G$ is simple. The curvature tensor admits exactly two independent solutions $k, u$ of equation (3) so that $\dim N_p = 2$. The Riemann tensor at $p$ takes the form

$$R_{abcd} = \alpha\, G_{ab} G_{cd}, \tag{6}$$

where $\alpha \in R$ and $G$ is simple bivector with blade orthogonal to $k$ and $u$.

**Class O**

The rank of the curvature matrix is 0 (so that $R_{abcd} = 0$) and $\dim N_p = 4$.

**Class A**

The Riemann tensor is said to be of class A at $p$ if it is not of class B, C, D or O. Here always $\dim N_p = 0$.

A study of the CCS for the classes A, B, C, D and O can be found in [5].

## 3. Main Results

Consider the Bianchi type IV space-times in the usual coordinate system $(t, x, y, z)$ (labeled by $(x^0, x^1, x^2, x^3)$, respectively) with line element [22]

$$ds^2 = -dt^2 + e^{-2z}\left[A(t)dx^2 + (z^2 A(t) + B(t))dy^2 + 2zA(t)dxdy\right] + C(t)dz^2 \qquad (7)$$

where $A(t), B(t)$ and $C(t)$ are no where zero functions of $t$. The above space-time admits three linearly independent Killing vector fields which are

$$\frac{\partial}{\partial x}, \quad \frac{\partial}{\partial y}, \quad (x-y)\frac{\partial}{\partial x} + y\frac{\partial}{\partial y} + \frac{\partial}{\partial z}. \qquad (8)$$

The non-zero independent components of the Riemann tensor are

$$R_{0101} = -\frac{1}{4A}\left[2A\ddot{A} - \dot{A}^2\right]e^{-2z} \equiv \alpha_1,$$

$$R_{0102} = -\frac{1}{4A}\left[2A\ddot{A} - \dot{A}^2\right]z\, e^{-2z} \equiv \alpha_7,$$

$$R_{0113} = \frac{1}{2C}\left[A\dot{C} - \dot{A}C\right]e^{-2z} \equiv \alpha_8,$$

$$R_{0123} = \frac{1}{4BC}\left[2\dot{A}BC - AC\dot{B} - AB\dot{C} + 2B(A\dot{C} - \dot{A}C)z\right]e^{-2z} \equiv \alpha_9,$$

$$R_{0202} = -\frac{1}{4AB}\left[2AB\ddot{B} - A\dot{B}^2 + (2A\ddot{A} - \dot{A}^2)Bz^2\right]e^{-2z} \equiv \alpha_2,$$

$$R_{0213} = \frac{1}{4C}\left[\dot{A}C - A\dot{C} + 2(A\dot{C} - \dot{A}C)z\right]e^{-2z} \equiv \alpha_{10},$$

$$R_{0223} = \frac{1}{4BC}\left[2B^2\dot{C} - 2B\dot{B}C + (3\dot{A}BC - A\dot{B}C - 2AB\dot{C})z + 2(AB\dot{C} - \dot{A}BC)z^2\right]e^{-2z} \equiv \alpha_{11},$$

$$R_{0303} = \frac{1}{4C}\left[\dot{C}^2 - 2C\ddot{C}\right] \equiv \alpha_3,$$

$$R_{1212} = \frac{1}{4C}\left[\dot{A}\dot{B}C + \dot{A}^2 - 4AB\right]e^{-4z} \equiv \alpha_4,$$

$$R_{1313} = \frac{1}{4B}\left[\dot{A}^2 - 4AB + \dot{A}B\dot{C}\right]e^{-2z} \equiv \alpha_5,$$

$$R_{0312} = \frac{1}{4B}\left[A\dot{B} - \dot{A}B\right]e^{-2z} \equiv \alpha_{12},$$

$$R_{1323} = \frac{1}{4B}\left[4AB - 4zAB + z\dot{A}B\dot{C} + zA^2\right]e^{-2z} \equiv \alpha_{13},$$

$$R_{2323} = \frac{1}{4B}\left[-3AB + 8zAB - 4z^2 AB - 4B^2 + z^2 \dot{A}B\dot{C} + B\dot{B}\dot{C} + A^2 z^2\right]e^{-2z} \equiv \alpha_6.$$

Writing the curvature tensor with components $R_{abcd}$ at $p$ as a $6 \times 6$ symmetric matrix

$$R_{abcd} = \begin{pmatrix} \alpha_1 & \alpha_7 & 0 & 0 & \alpha_8 & \alpha_9 \\ \alpha_7 & \alpha_2 & 0 & 0 & \alpha_{10} & \alpha_{11} \\ 0 & 0 & \alpha_3 & \alpha_{12} & 0 & 0 \\ 0 & 0 & \alpha_{12} & \alpha_4 & 0 & 0 \\ \alpha_8 & \alpha_{10} & 0 & 0 & \alpha_5 & \alpha_{13} \\ \alpha_9 & \alpha_{11} & 0 & 0 & \alpha_{13} & \alpha_6 \end{pmatrix} \qquad (9)$$

It is important to note that we will consider Riemann tensor components as $R^a{}_{bcd}$ for calculating CCS. We know from theorem [5] that when the rank of the $6 \times 6$ Riemann matrix is greater than three there exists no proper CCS. Therefore we are only interested in those cases when the rank of the $6 \times 6$ Riemann matrix is less than or equal to three. In general for any $6 \times 6$ symmetric matrix there exist total fourty one possibilities when the rank of the $6 \times 6$ symmetric matrix is less or equal to three, that is, twenty possibilities for rank three, fifteen possibilities for rank two and six possibilities for rank one. Suppose the rank of the $6 \times 6$ Riemann matrix is one. Then there is only one non-zero row or column in (9). If we set five rows or columns identically zero in (9) then there exist six possibilities when the rank of the $6 \times 6$ Riemann matrix is one. All these six possibilities give us contradiction. For example consider the case when the rank of the $6 \times 6$ Riemann matrix is one i.e. $\alpha_2 = \alpha_3 = \alpha_4 = \alpha_5 = \alpha_6 = \alpha_7 = \alpha_8 = \alpha_9 = \alpha_{10} = \alpha_{11} = \alpha_{12} = \alpha_{13} = 0$ and $\alpha_1 \neq 0$. Subsituting the above information back in equation (9) one has $\alpha_1 = 0$ which gives contradiction (here we assume that $\alpha_1 \neq 0$). So this case is not possible. Similarly if one proceeds further one finds that there exists only one possibility when the rank of the $6 \times 6$ Riemann matrix is three or less which is: $2A\ddot{A} - \dot{A}^2 = 0$, $A\dot{C} - \dot{A}C = 0$, $\dot{A}B - A\dot{B} = 0$, $2B\ddot{B} - \dot{B}^2 = 0$, $B\dot{C} - \dot{B}C = 0$, $2C\ddot{C} - \dot{C}^2 = 0$ and the rank of the $6 \times 6$ Riemann matrix is three.

In this case we have $B(t) = a A(t)$, $C(t) = b A(t)$ and $B(t) = (a/b) C(t)$, where $a, b \in R \setminus \{0\}$. Equations $2A\ddot{A} - \dot{A}^2 = 0$, $2B\ddot{B} - \dot{B}^2 = 0$ and $2C\ddot{C} - \dot{C}^2 = 0 \Rightarrow$ $A = (e_1 t + e_2)^2$, $B = (b_1 t + b_2)^2$ and $C = (f_1 t + f_2)^2$, where $e_1, e_2, b_1, b_2, f_1, f_2 \in R (e_1 \neq 0, b_1 \neq 0, f_1 \neq 0)$. It follows from the above calculation that $a e_1 = b_1$, $b e_1 = f_1$, $a e_2 = b_2$ and $b e_2 = f_2$. The sub case when $e_1 = 0$ (which implies $b_1 = 0$ and $f_1 = 0$) will be considered latter. Here, the rank of the $6 \times 6$ Riemann matrix is three and there exists a unique (up to a multiple) no where zero timelike vector field $t_a = t_{,a}$ solution of equation (3) and $t_{a;b} \neq 0$. The line element in this case takes the form

$$ds^2 = -dt^2 + (e_1 t + e_2)^2 \{e^{-2z}[dx^2 + (z^2 + a)dy^2 + 2zdxdy] + b\,dz^2\}. \tag{10}$$

Substituting the above information into CC equations (2) and after tedious and lengthy calculation we get

$$X^0_{,1} = X^0_{,2} = X^0_{,3} = 0, X^1_{,0} = X^1_{,3} = 0,$$
$$X^2_{,0} = X^2_{,3} = 0, X^3_{,0} = X^3_{,1} = X^3_{,2} = 0, \tag{11}$$

$$(1-2z)X^3 + z(X^1_{,1} + X^2_{,2}) + (a+z^2)X^2_{,1} + X^1_{,2} = 0, \tag{12}$$

$$(z - z^2 - a)X^3 + z\,X^1_{,2} + (a+z^2)X^2_{,2} = 0, \tag{13}$$

$$X^3 + 2(abd + z)X^3_{,3} + X^1_{,2} = 0, \tag{14}$$

$$(1-2z)X^3 - [a + (z-z^2)](X^1_{,1} - X^2_{,2} - 2X^3_{,3}) - (2z-1)X^1_{,2} = 0, \tag{15}$$

$$X^3 + 2[-abd + (z-1)]X^3_{,3} + X^1_{,2} - [a + (z-z^2)]X^2_{,1} = 0, \tag{16}$$

$$[2X^3_{,3} + X^1_{,1} - X^2_{,2}] + (abd + z)X^2_{,1} = 0, \tag{17}$$

$$X^3 - X^1_{,1} - zX^2_{,1} = 0, \tag{18}$$

$$(1-2z)X^3 + (a+z^2)X^2_{,1} + z\,X^1_{,1} + X^1_{,2} + zX^2_{,2} = 0, \tag{19}$$

where $d = \dfrac{4a - 4ae_1^2 b - 1}{4ab} \neq 0$. Equation (11) gives $X^0 = D(t)$, $X^1 = F^1(x,y)$, $X^2 = F^2(x,y)$ and $X^3 = F^3(z)$, where $D(t)$, $F^1(x,y)$, $F^2(x,y)$ and $F^3(z)$ are functions of integration. If one proceeds further one finds the solution of the above equations from (11) to (19)

$$X^0 = D(t), X^1 = c_1(x-y) + c_2, X^2 = c_1 y + c_3, X^3 = c_1, \tag{20}$$

where $D(t)$ is an arbitrary function of $t$ only and $c_1, c_2, c_3 \in R$. One can write the above equation (20) after subtracting the killing vector fields as

$$X = (D(t), 0, 0, 0). \tag{21}$$

Clearly CCS in this case form an infinite dimensional vector space.

Now consider the sub case when $e_1 = 0$. The above space-time (10) becomes

$$ds^2 = -dt^2 + e_2^2 \{e^{-2z}[dx^2 + (z^2 + a)dy^2 + 2zdxdy] + b\,dz^2\}. \tag{22}$$

The above space-time is 1+3 decomposable and belongs to curavture class C. In this sub case there exists a nowhere zero timelike vector field $t_a = t_{,a}$ such that $t_{a;b} = 0$. From the Ricci identity $R^a{}_{bcd} t_a = 0$. CCS in this case [5] are

$$X = D(t)\frac{\partial}{\partial t} + X', \qquad (23)$$

where $D(t)$ is an arbitrary function of $t$ only and $X'$ is a homothetic vector field in the induced geometry on each of the three dimensional submanifolds of constant $t$. The completion of this sub case needs to find a homothetic vector fields in the induced geometry of the submanifolds of constant $t$. The induced metric $g_{\alpha\beta}$ (where $\alpha, \beta = 1, 2, 3$) with non zero components is given by

$$g_{11} = e_2^2 e^{-2z}, \; g_{12} = e_2^2 z e^{-2z}, g_{22} = e_2^2 (z^2 + a) e^{-2z}, g_{33} = e_2^2 b. \qquad (24)$$

A vector field $X'$ is a homothetic vector field if it satisfies $L_{X'} g_{\alpha\beta} = 2\phi g_{\alpha\beta}$, for all $\alpha, \beta = 1, 2, 3,$ where $\phi \in R$. One can expand the homothetic equation and using (24) to get

$$X^3 - X^1_{,1} - z X^2_{,1} = -\phi, \qquad (25)$$

$$(1 - 2z)X^3 + zX^1_{,1} + (z^2 + a)X^2_{,1} + X^1_{,2} + zX^2_{,2} = 2\phi z, \qquad (26)$$

$$bX^3_{,1} + e^{-2z} X^1_{,3} + z e^{-2z} X^2_{,3} = 0, \qquad (27)$$

$$(z - z^2 - a)X^3 + z X^1_{,2} + (z^2 + a)X^2_{,2} = \phi(z^2 + a), \qquad (28)$$

$$b X^3_{,2} + z e^{-2z} X^1_{,3} + (z^2 + a)e^{-2z} X^2_{,3} = 0, \qquad (29)$$

$$X^3_{,3} = \phi. \qquad (30)$$

Equation (30) gives $X^3 = \phi z + E^1(x, y),$ where $E^1(x, y)$ is a function of integration. Multiply equation (27) with $z$ and subtracting from equation (29) and using the above value of $X^3$ and upon integration we get

$$X^2 = \frac{b}{a}\int (zE^1_x(x, y) - E^1_y(x, y))e^{2z} dz + E^2(x, y),$$

where $E^2(x, y)$ is a function of integration. Substituting the above information in equation (27) and upon integration one has

$$X^1 = -\frac{b}{6a}[3a + 3z^2 + 2z^3] E^1_x(x, y) e^{2z} - [3z + 3z^2] E^1_y(x, y) + E^3(x, y),$$

where $E^3(x,y)$ is a function of integration. In order to determine homothetic vector field we need to calculate $E^1(x,y)$, $E^2(x,y)$ and $E^3(x,y)$. If one proceeds further one finds that $\phi = 0$ which implies that no proper homothetic vector field exists in the induced geometry of the submanifolds of constant $t$. Hence homothetic vector fields are Killing vector fields which are

$$X^1 = c_1(x-y) + c_2, \quad X^2 = c_1 y + c_3, \quad X^3 = c_1, \tag{31}$$

where $c_1, c_2, c_3 \in R$. CCS in this case are (using equation (31) in (23)) given in equation (20). Clearly CCS in this case form an infinite dimensional vector space.

**SUMMARY**

In this paper an attempt is made to explore all the possibilities when the Bianchi type IV space-times admit proper CCS. An approach is adopted to study proper CCS of the above space-times by using the rank of the $6 \times 6$ Riemann matrix and also using the theorem given in [5], which suggested where proper curvature collineations exist. From the above approach we obtain only one case when the above space-times (10) admit proper CCS which forms an infinite dimensional vector space. The subcase of the above case is also discuss when the space-times become static and admits proper CCS which form an infinite dimensional vector space. This is the space-times (22).

# References

[1] G. H. Katzin, J. Levine and W. R. Davis, J. Math. Phys. **10** (1969) 617.

[2] G. Shabbir, Grav. Cosmol., **9** (2003) 139.

[3] G. Shabbir, Nuovo Cimento B, **118** (2003) 41.

[4] G. S. Hall and G. Shabbir, Classical Quantum Gravity, **18** (2001) 907.

[5] G. S. Hall and J. da. Costa, J. Math. Phys. **32** (1991) 2854.

[6] G. S. Hall, Symmetries and Curvature Structure in General Relativity, World Scientific, 2004.

[7] G. S. Hall, Gen. Rel. Grav. **15** (1983) 581.

[8] A. H. Bokhari, A. Qadir, M. S. Ahmed and M. Asghar, J. Math. Phys. **38** (1997) 3639.

[9] R. A. Tello-Llanos, Gen. Rel. Grav. **20** (1988) 765.

[10] J. Carot and J. da. Costa, Gen. Rel. Grav. **23** (1991) 1057.

[11] G. S. Hall, Classical Quantum Gravity, **23** (2006) 1485.

[12] G. S. Hall and Lucy MacNay, Classical Quantum Gravity, **22** (2005) 5191.

[13] A. H. Bokhari, M. Asghar, M. S. Ahmed, K. Rashid and G. Shabbir, Nuovo Cimento B, **113** (1998) 349.

[14] G. Shabbir, A. H. Bokhari and A. R. Kashif, Nuovo Cimento B, **118** (2003) 873.

[15] G. S. Hall and G. Shabbir, Grav. Cosmol., **9** (2003) 134.

[16] G. Shabbir, Nuovo Cimento B, **119** (2004) 433.

[17] G. Shabbir, Nuovo Cimento B, **121** (2006) 319.

[18] G. Shabbir and Abu Bakar Mehmood, Modern Physics Letters A, **22** (2007) 807.

[19] G. Shabbir and M. Ramzan, International journal of Modern Physics A, **23** (2008) 749.

[20] G. Shabbir, Amjad Ali and M. Ramzan, to appear in Gravitation and Cosmology in 2010.

[21] G. Shabbir, Applied and Computational Mathematics An International journal, **2** (2003) 65.

[22] H. Stephani, D. Kramer, M. A. H. MacCallum, C. Hoenselears and E. Herlt, *Exact Solutions of Einstein's Field Equations*, Cambridge University Press, 2003.